\documentclass[prd,aps,a4,twocolumn,superscriptaddress,preprintnumbers,nofootinbib]{revtex4-1}

\pdfoutput=1

\usepackage{pslatex}
\usepackage[pdftex]{graphicx}
\usepackage{psfrag}
\usepackage{epsfig}
\usepackage{color}
\usepackage{cancel}
\usepackage{slashed}
\usepackage{amssymb}
\usepackage{amsmath}
\usepackage{hyperref}
\usepackage{enumerate}
\usepackage{multirow}
\usepackage{natbib}
\usepackage[normalem]{ulem}

\definecolor{dgreen}{rgb}{0,0.5,0} 

\begin{document}
\title{Sensitivity of direct detection
  experiments to neutrino magnetic dipole moments}%
\author{D. Aristizabal Sierra}
\email{daristizabal@ulg.ac.be}
\affiliation{Universidad T\'ecnica Federico Santa Mar\'{i}a -
  Departamento de F\'{i}sica\\ Casilla 110-V, Avda. Espa\~na 1680,
  Valpara\'{i}so, Chile}
\affiliation{IFPA, Dep. AGO, Universit\'e de Li\`ege, Bat B5, Sart Tilman B-4000 Li\`ege 1, Belgium}
\author{R. Branada}
\email{rocio.branada@sansano.usm.cl}
\affiliation{Universidad T\'ecnica Federico Santa Mar\'{i}a -
    Departamento de F\'{i}sica\\ Casilla 110-V, Avda. Espa\~na 1680,
    Valpara\'{i}so, Chile}
\author{O. G. Miranda}
\email{omr@fis.cinvestav.mx}
\affiliation{Departamento de
  F\'{i}sica, Centro de Investigaci\'on y de Estudios Avanzados del
  IPN, Apartado Postal 14-740 07000 Mexico, Distrito Federal, Mexico}
\author{G. Sanchez Garcia}
\email{gsanchez@fis.cinvestav.mx}
\affiliation{Departamento de F\'{i}sica, Centro de Investigaci\'on y
  de Estudios Avanzados del IPN, Apartado Postal 14-740 07000 Mexico,
  Distrito Federal, Mexico}
\begin{abstract}
  With large active volume sizes dark matter direct detection
  experiments are sensitive to solar neutrino fluxes. Nuclear recoil
  signals are induced by $^8$B neutrinos, while electron recoils are
  mainly generated by the pp flux. Measurements of both processes
  offer an opportunity to test neutrino properties at low thresholds
  with fairly low backgrounds. In this paper we study the sensitivity
  of these experiments to neutrino magnetic dipole moments assuming 1,
  10 and 40 tonne active volumes (representative of XENON1T, XENONnT
  and DARWIN), 0.3 keV and 1 keV thresholds. We show that with nuclear
  recoil measurements alone a 40 tonne detector could be as
  competitive as Borexino, TEXONO and GEMMA, with sensitivities of
  order $8.0\times 10^{-11}\,\mu_B$ at the $90\%$ CL after one year of
  data taking. Electron recoil measurements will increase
  sensitivities way below these values allowing to test regions not
  excluded by astrophysical arguments. Using electron recoil data and
  depending on performance, the same detector will be able to explore
  values down to $4.0\times 10^{-12}\mu_B$ at the $90\%$ CL in one
  year of data taking. By assuming a 200-tonne liquid xenon detector
  operating during 10 years, we conclude that sensitivities in this
  type of detectors will be of order $10^{-12}\,\mu_B$. Reducing
  statistical uncertainties may enable improving sensitivities below
  these values.
\end{abstract}
\maketitle
\section{Introduction}
\label{sec:intro}
Dark matter (DM) direct detection experiments are already sensitive to
solar neutrinos. In its latest data sets, XENON1T has reported signals
in both coherent elastic neutrino-nucleus scattering (CE$\nu$NS) and
neutrino-electron elastic
scattering~\cite{Aprile:2019xxb,Aprile:2020tmw}. It is natural to
expect that with increasing active volumes and exposures,
XENONnT~\cite{Aprile:2015uzo}, LZ~\cite{Akerib:2018lyp}, and
DARWIN~\cite{Aalbers:2016jon} will provide larger statistics in both
channels. Threfore, their results will enable precise measurements of
neutrino properties complementing those coming from present and
near-future dedicated neutrino experiments (see,
e.g.~\cite{Deniz:2009mu,An:2015jdp,Acciarri:2015uup,Strauss:2017cuu,Hakenmuller:2019ecb,Aguilar-Arevalo:2019jlr}). The
opportunities these data offer include---but are not limited to---
studies of new interactions in the neutrino sector by means of light
vector and scalar mediators, neutrino non-standard interactions, and
neutrino electromagnetic properties
\cite{Baudis:2013qla,Cerdeno:2016sfi,Dutta:2017nht,AristizabalSierra:2017joc,Gonzalez-Garcia:2018dep,AristizabalSierra:2019ykk,Papoulias:2018uzy,Hsieh:2019hug}. They
will also provide a playground for precise measurements of solar
neutrino fluxes, including those from the solar CNO
cycle~\cite{Newstead:2018muu}, as well as for tests of solar models
and solar neutrino matter effects~\cite{Aalbers:2020gsn}.

With precise discrimination, measurements of electron or nuclear
recoils alone can determine the presence of new physics.  That could
be the case---for instance---of the recent electron excess reported by
the XENON1T collaboration \cite{Aprile:2020tmw}, if indeed new physics
is responsible for such a signal. Ideally, a physical explanation of
an electron excess would produce a particular signature in the
corresponding nuclear channel. However, an observation of a signal
e.g. in electron recoils does not necessarily implies an observation
of nuclear recoils. The main reason being the energy thresholds
involved. For $\sim 0.1\,$keV thresholds, electron recoils are driven
by pp neutrino fluxes, while nuclear recoils by $^8$B neutrinos. In
$\text{cm}^{-2}\text{s}^{-1}$ units, these fluxes differ by about four
orders of magnitude~\cite{Bahcall:2004pz}. Thus, unless the new
physics effects are way more pronounced in the $\nu-N$ cross section,
one expects the electron recoil signal to be more prominent (possibly
only observable in that channel). An extreme case of such scenario
would be a nucleon-phobic light vector mediator, in which case only an
excess in electron recoils could be observed.

However, there are other cases in which either the electron or nuclear
recoil signal can come along with a signal in the other
channel. Arguably the most remarkable example in this case is given by
photon exchange. As soon as the new physics couples to photons,
depending on the size of the new physics couplings, there will be new
contributions to electron and nuclear recoil signals \footnote{Note
  that even in the most extreme case, nucleon- or lepto-phobic
  interactions, reconstruction of the signal will require electron and
  nuclear recoil measurements. For neutrino nuclear recoil traces of
  lepto-phobic scenarios see
  ~\cite{AristizabalSierra:2019ufd,AristizabalSierra:2019ykk}.}.  In
the case of neutrino detection, any of its electromagnetic couplings
will lead to both electron and nuclear recoils, again depending on
their size both processes could be observed. This discussion is of
course not only related to neutrinos, a typical example involving DM
is given by dark photon portals in which kinetic mixing allows
coupling of the dark and visible (SM) sectors
\cite{Okun:1982xi,Holdom:1985ag}.

Motivated by the latest XENON1T result~\cite{Aprile:2020tmw}, in this
paper we study the extent at which neutrino electromagnetic properties
can be tested at XENON1T, XENONnT and DARWIN using combined electron
and nuclear recoil measurements. We consider neutrino magnetic dipole
moments and determine the discovery reach under simplified detector
and signal assumptions. This includes one, ten and forty tonne active
volumes, $100\%$ detector efficiency and $0.3\,\text{keV}$ and
$1\,\text{keV}$ energy thresholds. The former motivated by
Ref.~\cite{Lenardo:2019fcn}, while the latter determined by future
detector performances
\cite{Aprile:2015uzo,Akerib:2018lyp,Aalbers:2016jon}. In all cases our
\textit{toy experiments} correspond to the SM electron and nuclear
recoil spectra (measured in events/tonne/year/keV). For CE$\nu$NS we
assume two background hypotheses, $68\%$ and $25\%$ of the signal
rate. While for $\nu-e$ elastic scattering we instead use expected
backgrounds at XENON1T, XENONnT and DARWIN as given in
Refs. \cite{Aprile:2020tmw,Aprile:2020vtw,Aalbers:2020gsn}. Needless
to say, these assumptions---in particular for CE$\nu$NS---are just
representative of how the actual detectors performances and data sets
will look like, but allow us to visualize how competitive these
detectors will be when compared to neutrino dedicated experiments.

The remainder of this paper is organized as follows. In
Sec. \ref{sec:CEvNS-NMM} we shortly discuss CE$\nu$NS, $\nu-e$ elastic
scattering as well as neutrino magnetic dipole moments and their
corresponding differential cross sections. In
Sec. \ref{sec:param-space-analysis} we describe our statistical
analysis and present our results. Finally in
Sec. \ref{sec:conclusions} we summarize and present our conclusions.
\section{CE$\nu$NS, $\nu-e$ elastic scattering and neutrino magnetic dipole moments}
\label{sec:CEvNS-NMM}
In the SM, CE$\nu$NS is a neutral current process in which the
neutrino energy involved, $E_\nu\lesssim 100\,$MeV, is such that the
transferred momentum implies the individual nucleon amplitudes sum up
coherently. This results in an approximately overall enhancement of
the cross section, determined by $N^2$, where $N$ refers to the number
of neutrons of the target nuclei
\cite{Freedman:1973yd,Freedman:1977xn}. Given the constraints over the
neutrino energy probe, possible neutrinos that can induce the process
are limited to neutrinos produced in pion decay-at-rest, reactor and
solar neutrinos. Other possible sources include sub-GeV atmospheric
neutrinos, diffuse supernova (SN) background (DSNB) neutrinos and a SN
burst. However, their detection is less certain. Sub-GeV atmospheric
neutrinos and DSNB fluxes are small, and so very large exposures are
required for their detection~\cite{Newstead:2020fie}. Observation of a
SN burst is not guaranteed, although it is expected to happen at
certain point~\cite{Lang:2016zhv}.

The differential cross section for this process involves a
zero-transferred momentum component and a nuclear form factor that
accounts for nuclear structure. It is given by~\cite{Freedman:1973yd}
\begin{equation}
  \label{eq:CEvNS_diff_xsec}
  \frac{d\sigma_{\nu-N}}{dE_r}=\frac{G_F^2}{2\pi}m_Ng_N^2
  \left(
    2 - \frac{m_NE_r}{E_\nu^2}
  \right)\,F^2(E_r)\ .
\end{equation}
Here $g_N=(A-Z)g_N^n+Zg_V^p$, where $g_N^n=2g_N^d+g_N^u$ and
$g_N^p=g_N^d+2g_N^u$ with the quark electroweak couplings given by
$g_N^d=-1/2+2/3\sin^2\theta_W$ and $g_N^u=1/2-4/3\sin^2\theta_W$. For
the weak mixing angle we use $\sin^2\theta_W=0.23122$
\cite{1674-1137-40-10-100001}. For the nuclear form factor we adopt
the Helm parametrization and assume the the same root-mean-square
radii for the proton and neutron distributions. Assuming otherwise
requires weighting the neutron and proton contributions with
independent form factors~\cite{AristizabalSierra:2019zmy}. Note that
for the energies we are interested in the form factor plays a somewhat
minor role.

Solar neutrinos are subject to neutrino flavor conversion, which
depending on the process of the pp chain they originate from can be
matter enhanced. Assuming the two-flavor approximation (mass dominance
limit $\Delta m_{13}^2\to \infty$), two neutrino flavors reach the
detector. One mainly an electron flavor, $\nu_e$, with a muon
contamination suppressed by the reactor mixing angle. And a second
one, $\nu_a$, that is a superposition of muon and tau flavors with the
admixture controlled by the atmospheric mixing
angle. Neutrino-electron scattering induced by solar neutrinos
receives therefore contributions from neutral and charged
current. Neutral from $\nu_e-e$ and $\nu_a-e$ interactions, while
charged from $\nu_e-e$ alone. The differential cross section
reads~\cite{Vogel:1989iv}
\begin{align}
  \label{eq:nu-e-xsec}
  \frac{d\sigma_{\nu-e}}{dE_r}=\frac{G_F^2m_e}{2\pi}
  &\left[
    (g_V+g_A)^2 
    + (g_V-g_A)^2
    \left(
      1 - \frac{E_r}{E_\nu}
    \right)^2
    \right .
    \nonumber\\
  & \left . 
    (g_A^2 - g_V^2)\frac{m_eE_r}{E_\nu^2}
    \right]\ ,
\end{align}
with the vector and axial couplings given by
\begin{align}
  \label{eq:nu-e-couplings}
  g_V=2\sin^2\theta_W \pm \frac{1}{2}\ ,\qquad
  g_A= \pm \frac{1}{2}\ .
\end{align}
Here `$+$' holds for $\nu_e$, while `$-$' for $\nu_a$.
\subsection{Neutrino magnetic/electric dipole moments
and cross sections}
\label{sec:nmm}
Possible neutrino electromagnetic couplings are determined by the
neutrino electromagnetic current, which decomposed in terms of
electromagnetic form factors leads to four diagonal independent
couplings in the zero-transferred momentum limit: electric charge,
magnetic dipole
moment~\cite{Fujikawa:1980yx,Schechter:1981hw,pal:1982rm,Kayser:1982br,Nieves:1981zt,Shrock:1982sc},
electric dipole moment and anapole
moment~\cite{Kayser:1982br}. Depending on whether neutrinos are Dirac
or Majorana and on whether CP and CPT are exact symmetries in the new
physics sector, some of these couplings may
vanish~\cite{Kayser:1982br,Nieves:1981zt}.  They are subject to a
variety of limits from laboratory experiments~\cite{Canas:2015yoa}
that include PVLAS~\cite{DellaValle:2015xxa}, neutron $\beta$
decay~\cite{Foot:1989fh}, TRISTAN, LEP and
CHARM-II~\cite{Hirsch:2002uv}, MUNU~\cite{Daraktchieva:2005kn},
Super-Kamiokande~\cite{Liu:2004ny}, TEXONO~\cite{Deniz:2009mu},
GEMMA~\cite{Beda:2010hk} and Borexino Phase-II
\cite{Borexino:2017fbd}. They are constrained by cosmology and
astrophysical observations as well, including primordial
nucleosynthesis~\cite{Grifols:1986ed}, neutrino star
turning~\cite{Studenikin:2012vi}, supernova and stellar
cooling~\cite{Barbiellini:1987zz,Grifols:1989vi,Raffelt:1990yz,Raffelt:1990pj}.
For an extensive review on this constraints see
Ref.~\cite{Giunti:2014ixa} (see Ref. \cite{Brdar:2020quo} as well
for constraints involving order MeV right-handed neutrinos).

At the effective level the magnetic/dipole couplings can be written
according to
\begin{align}
  \label{eq:eff-couplings-nmm}
  -\mathcal{L}^D&=\frac{1}{2}\overline{\nu_{L_i}}\sigma_{\mu\nu}
                  \left(\mu_{ij}^D + \gamma_5\epsilon_{ij}^D\right)\nu_{R_j}\  
                  + \text{H. c.} ,
                  \nonumber\\
  -\mathcal{L}^M&=\frac{1}{2}\overline{\nu_{L_i}^c}\sigma_{\mu\nu}
                  \left(\mu_{ij}^M 
                  + \gamma_5\epsilon_{ij}^M\right)\nu_{L_j}\ + \text{H. c.}
\end{align}
Note that electron recoil experiments cannot differentiate between
Dirac or Majorana couplings, nor between magnetic/electric moments or
transitions. In the mass eigenstate basis, processes induced by
interactions (\ref{eq:eff-couplings-nmm}) are sensitive to the
effective parameter \cite{Beacom:1999wx}
\begin{equation}
  \label{eq:eff-parameter-nmm}
  \mu_{\nu_\text{eff}}^2=
  \sum_j\left|\sum_k A_k(E_\nu,L)(\mu_{kj}-i\epsilon_{kj})\right|^2\ ,
\end{equation}
where $A_k(E_\nu,L)$ refers to the amplitude of the $k$-th massive
neutrino state at detection point. The effective coupling takes
different forms depending on whether neutrinos are Dirac or Majorana
as well as on the flavor scheme adopted (see
e.g. Ref.~\cite{Canas:2015yoa}). We will therefore use this effective
coupling in our calculations, since it can be useful from the
phenomenological point of view. As a simple illustration of this
point, we can consider the case of diagonal couplings $\mu_e$ and
$\mu_a$. In the two-flavor approximation and for oscillation
parameters as required by the LMA-MSW solution the effective coupling
takes the form \cite{Borexino:2017fbd}
\begin{equation}
  \label{eq:msw-lma-sol}
  \mu_{\nu_\text{eff}}^2\simeq P_{ee}\mu_e^2 + (1-P_{ee})\mu_a^2\ .
\end{equation}
Here $P_{ee}$ is the electron neutrino survival probability (see
discussion in next section) and we have used $\sin^4\theta_{13}\ll
1$.
The analysis of neutrino magnetic interactions is therefore a
multiparameter problem that can be reduced to a single parameter
problem with the aid of (\ref{eq:eff-parameter-nmm}). Using this
parametrization, the $\nu-e$ differential cross section
reads~\cite{Vogel:1989iv}
\begin{equation}
  \label{eq:nu-electron-nmm-xsec}
  \frac{d\sigma_\nu}{dE_r}=\pi\alpha\,\frac{\mu_{\nu_\text{eff}}^2}{m_e^2}
  \left(
    \frac{E_\nu-E_r}{E_r}
  \right)\ ,
\end{equation}
where $\mu_{\nu_\text{eff}}$ has been normalized to the Bohr magneton.
For CE$\nu$NS the differential cross section has the same structure
but comes along with the number of target protons squared $Z^2$ and a
nuclear form factor. Because of the Coulomb divergence the cross
section is forward peaked, a behavior that becomes rather pronounced
at low recoil energies. The most salient feature of neutrino magnetic
moment interactions is thus spectral distortions.

\begin{table}
  \centering
  \begin{tabular}{|c|c|}\hline
    \textbf{Component} & \textbf{Kinematic limit} [keV]\\\hline\hline
    pp & $2.64\times 10^2$\\\hline
    $^7$Be ($E_\nu=0.3\,$MeV) & $2.31\times 10^2$ \\\hline
    $^7$Be ($E_\nu=0.8\,$MeV) & $6.64\times 10^2$ \\\hline
    pep & $1.18\times 10^3$\\\hline
    hep & $1.85\times 10^4$\\\hline
    $^8$B & $1.63\times 10^4$ $(4.48)$\\\hline
    $^{13}$N & $9.88\times 10^2$\\\hline
    $^{15}$O & $1.51\times 10^3$\\\hline
    $^{17}$F & $1.52\times 10^3$\\\hline
  \end{tabular}
  \caption{Kinematic recoil energy limit for the different neutrino 
    components of the solar pp and CNO cycles in neutrino-electron
    scattering. Included as well in parenthesis is the  kinematic limit 
    for $^8$B in CE$\nu$NS. The values displayed follow from the BS05 
    standard solar model~\cite{Bahcall:2004pz}.}
  \label{tab:kinematic-upper-limits}
\end{table}
\begin{figure*}
  \centering
  \includegraphics[scale=0.37]{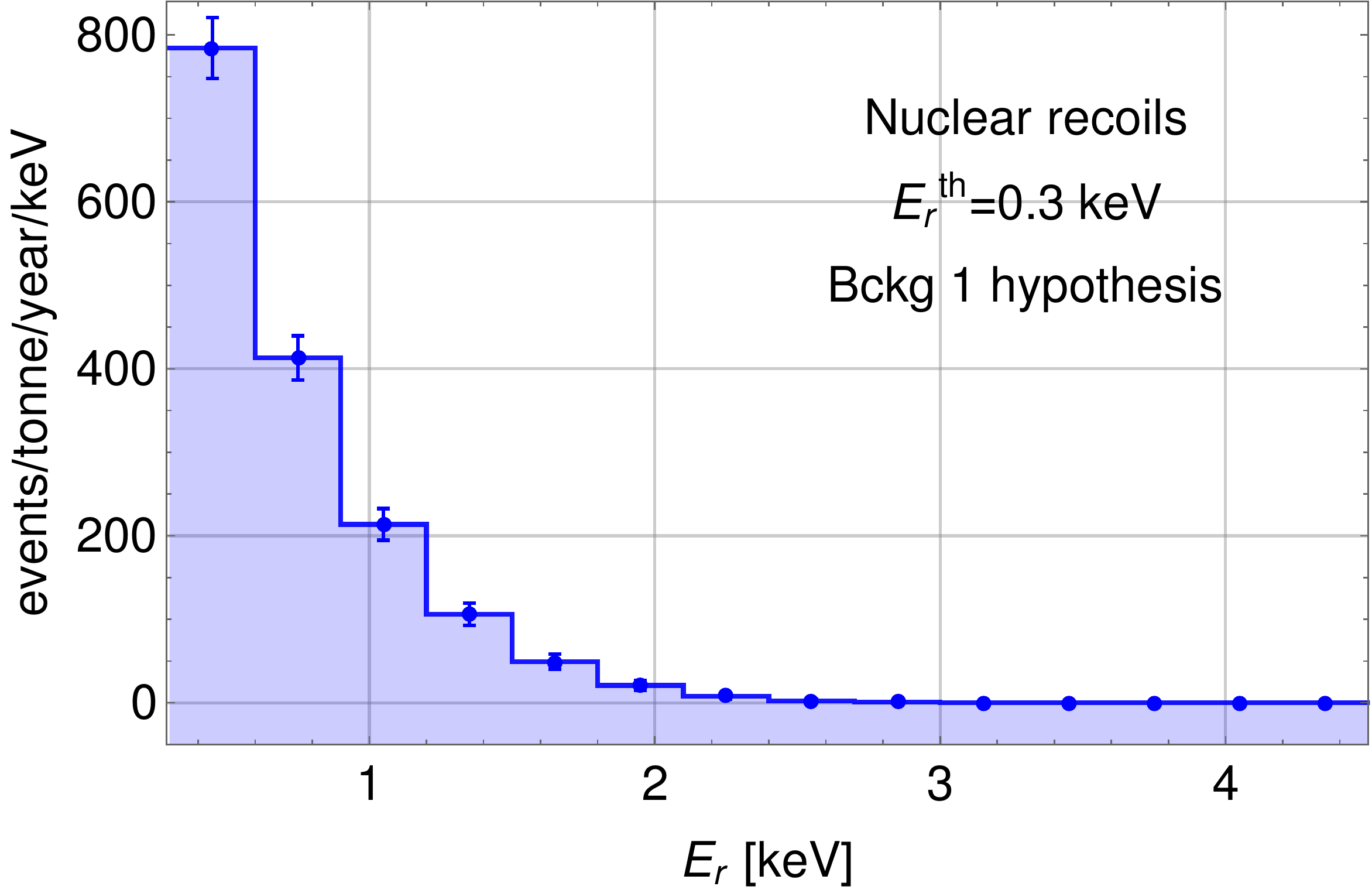}
  \includegraphics[scale=0.37]{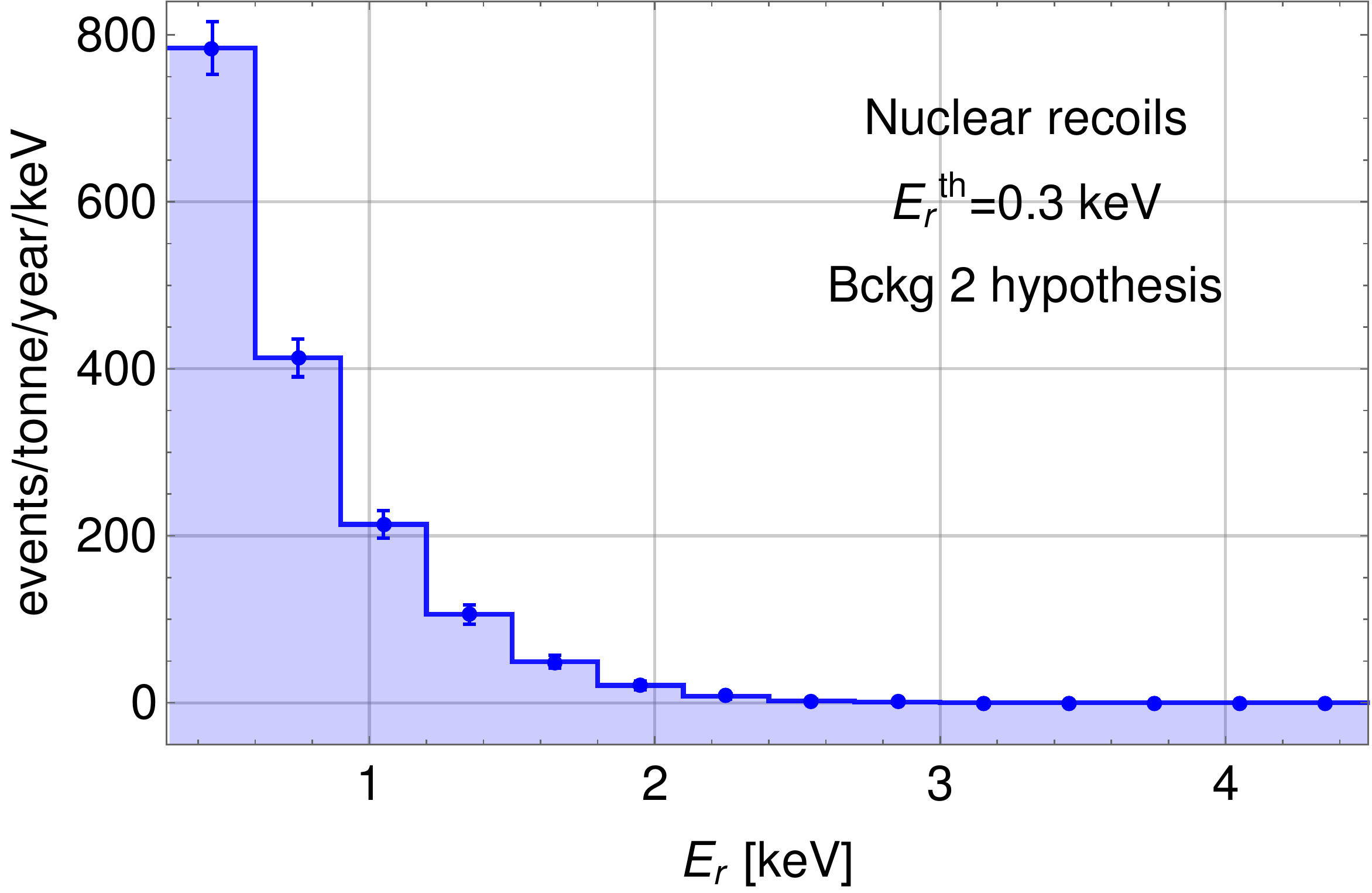}
  \includegraphics[scale=0.37]{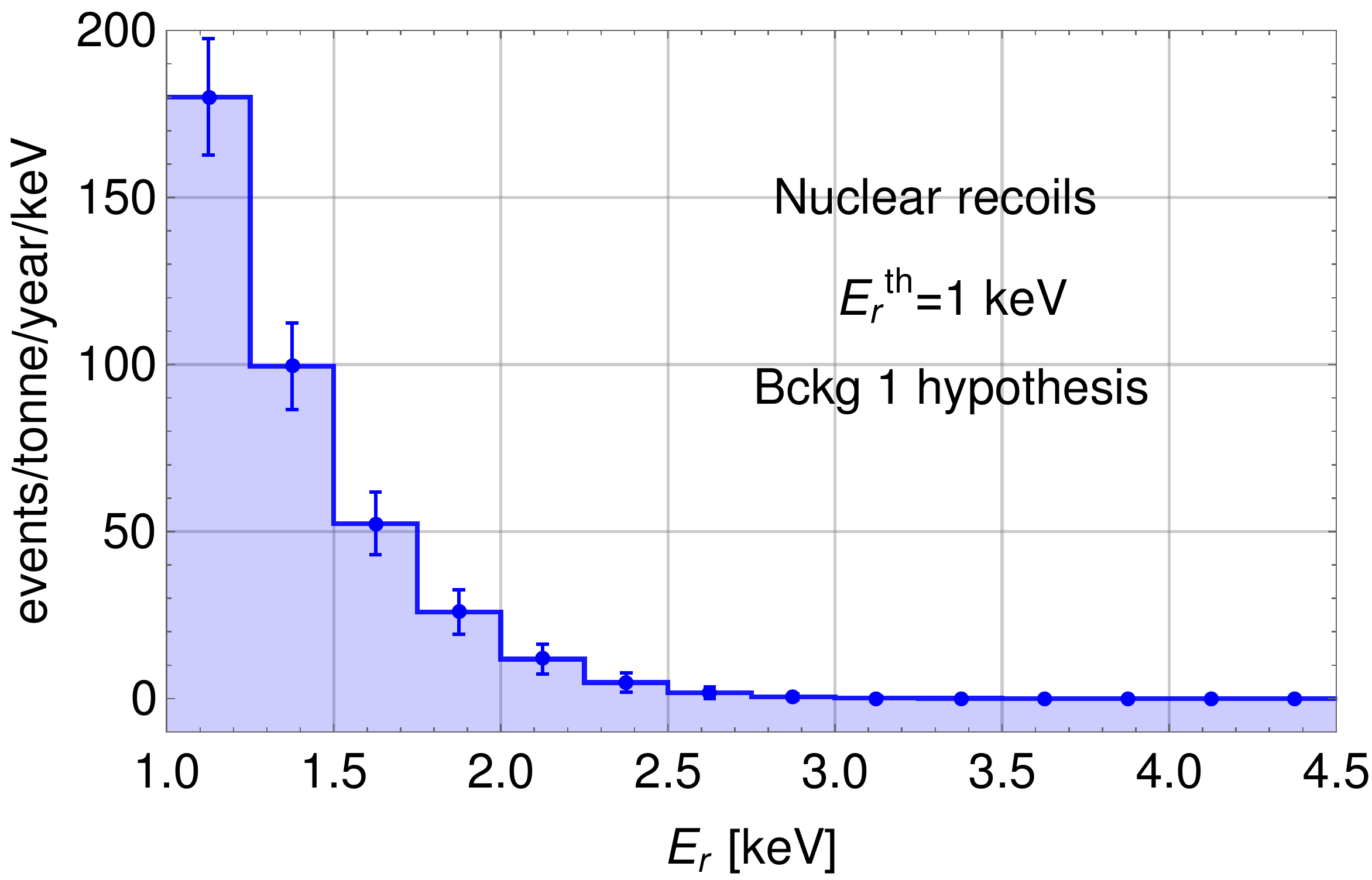}
  \includegraphics[scale=0.37]{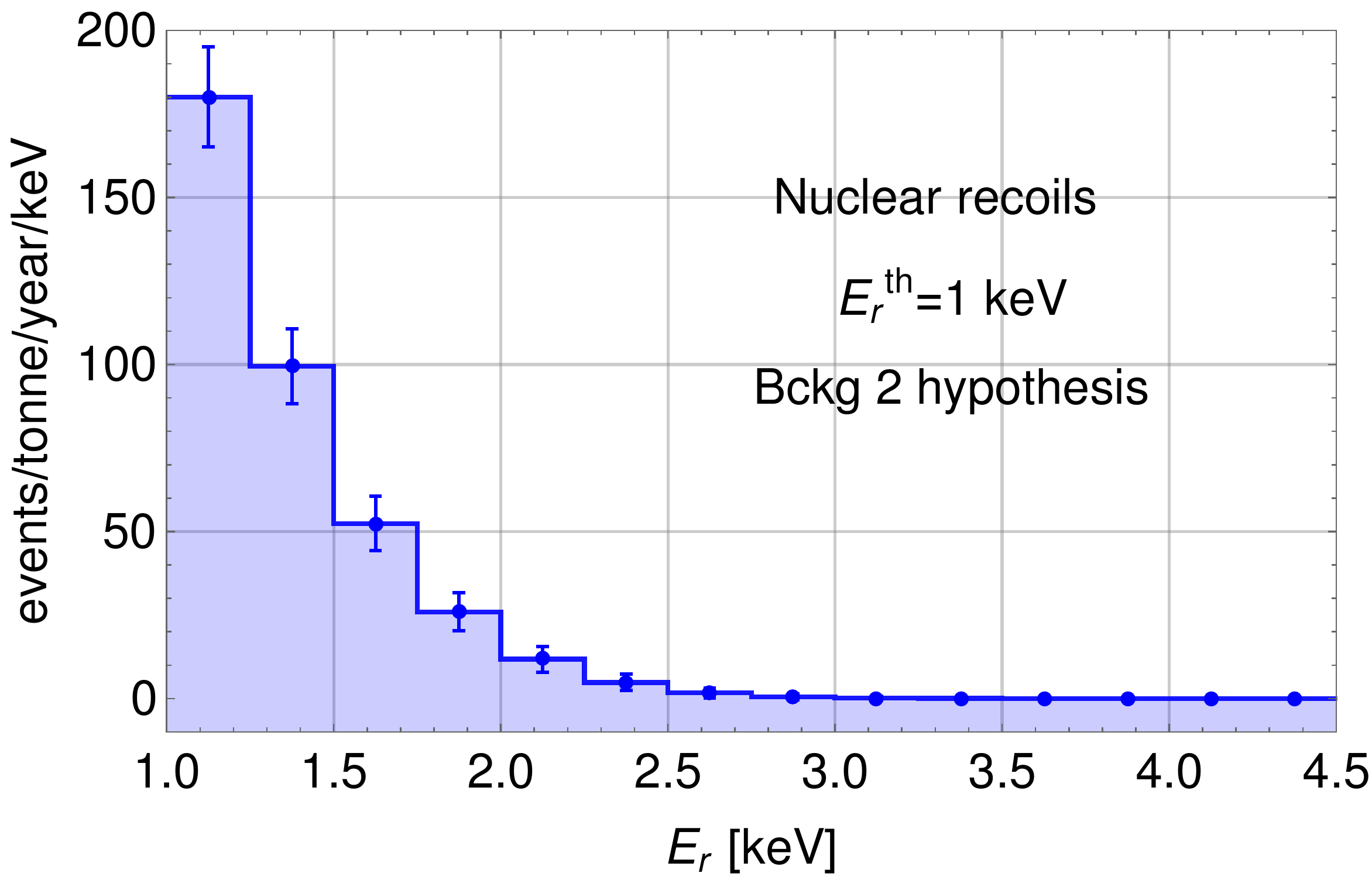}
  \caption{\textbf{Top graphs}: CE$\nu$NS toy experiment signals for a
    $0.3\,$keV threshold with a background amounting to $68\%$ (left)
    and $25\%$ (right) of the signal rate, Bckg 1 and Bckg 2
    hypotheses respectively. The result assumes 100\% detector
    efficiency and one tonne-year exposure. \textbf{Bottom graphs}:
    Same as the top graphs but for $1\,$keV threshold instead. For the
    CE$\nu$NS analysis, these are the ``experimental'' signals we have
    assumed. As can be seen, results are rather sensitive to threshold
    choices.}
  \label{fig:signals_CEvNS}
\end{figure*}
For all the cross sections we have discussed maximum recoil energies
are written as
\begin{align}
  \label{eq:Ermax}
  \text{Neutrino-electron}:\quad E_r^\text{max}&=\frac{2E_\nu^2}{m_e+2E_\nu}\ ,
                  \nonumber\\
  \text{Neutrino-nucleus}:\quad E_r^\text{max}&=\frac{2E_\nu^2}{m_N+2E_\nu}\simeq \frac{2E_\nu^2}{m_N}\ ,
\end{align}
with the approximation being fairly good for all isotopes of interest,
particularly for xenon.
\subsection{Recoil spectra}
\label{sec:recoil-spectrum}
The recoil spectrum for nuclear and electron recoils proceeds from a
convolution of neutrino fluxes and neutrino cross sections.  In the
case of nuclear recoils, they will be sensitive to all neutrino
flavors on equal footing. For electron recoils the situation is
different since electron neutrinos are subject as well to charged
current processes, while the other flavors do not. Fluxes, therefore,
should be weighted by the neutrino oscillation survival probability
$P_{ee}$, which proceeds from an average over neutrino trajectory,
including all neutrino fluxes $d\Phi/dE_\nu$ (pp and CNO cycles) and
involving neutrino production distributions as predicted by the
standard solar model (see e.g.~\cite{AristizabalSierra:2017joc}). For
its calculation we have employed those given by the BS05
model~\cite{Bahcall:2004pz} and the neutrino oscillation parameters
best-fit-point values in \cite{deSalas:2017kay}.  Inclusion of
neutrino magnetic moments can involve neutrino oscillation
probabilities too, depending on whether the new couplings are or not
flavor dependent.

For thresholds above $\sim 0.1\,$keV CE$\nu$NS is sensitive only to
the $^8$B neutrino flux (the hep flux is too suppressed to give a
sizable signal). Neutrino-electron elastic scattering instead is
sensitive to all solar neutrino fluxes, and so it is dominated by pp
neutrinos. Contribution form other fluxes is small, with the main
contribution given by the $0.86\,$MeV $^7$Be line (see
e.g.~\cite{AristizabalSierra:2020edu}). We write then the recoil
spectra as follows
\begin{align}
  \label{eq:rec-spectra}
  \frac{dR_{\nu-N}}{dE_r}=&
  N_N\int_{E_\text{min}}^{E_\text{max}}\,\frac{d\Phi_{^8\text{B}}}{dE_\nu}
  \left[
   \frac{d\sigma_{\nu-N}}{dE_r} + \frac{d\sigma_\nu}{dE_r}
  \right]dE_\nu\ ,
  \nonumber\\
  \frac{dR_{\nu-e}}{dE_r}=&
  N_e\int_{E_\text{min}}^{E_\text{max}}\,\sum_\alpha\frac{d\Phi_\alpha}{dE_\nu}
  \left[P_{ee}\frac{d\sigma_{\nu-e}}{dE_r}
                           \right .
                           \nonumber\\
  &\left . +(1-P_{ee})\frac{d\sigma_{\nu_a}}{dE_r} + \frac{d\sigma_\nu}{dE_r}
  \right]dE_\nu\ .
\end{align}
Here $N_N$ and $N_e$ refer to the number of nuclei and electrons in
the detector, $E_\nu^\text{min}=\sqrt{m_NE_r/2}$ for CE$\nu$NS and
$E_\nu^\text{min}=[E_r+(E_r^2+2E_rm_e)^{1/2}]/2$ for $\nu-e$
scattering. Index $\alpha$ runs over pp, $^8$B, hep, $^7$Be, pep
$^{13}$N, $^{15}$O and $^{17}$F. $E_\nu^\text{max}$ is determined by
the kinematic tail of the corresponding flux as displayed in
Tab.~\ref{tab:kinematic-upper-limits}.
\section{Sensitivity to neutrino magnetic moments}
\label{sec:param-space-analysis}
Xenon multi-ton scale DM detectors rely on photon (scintillation) and
electron (ionization) signals~\cite{Aprile:2015uzo,Aalbers:2016jon}.
Photons are detected through photosensors that produce a prompt S1
signal. Electrons, instead, are drifted upwards with the aid of an
electric field, resulting in a delayed S2 signal. S1 and S2 signals in
turn allow the reconstruction of the radial position and depth of a
given interaction, together with the energy reconstruction of an
event. Their ratio, S2/S1, provides a way to descriminate between
electron and nuclear recoils. Moreover, the dual-phase technology,
allows to get more information on the S2 signal improving the
resolution power of these detectors.  The combination of these
features provides a powerful tool for event selection over background
and this will eventually enable the reconstruction of new physics
signals, if any, through the discrimination of electronic and nuclear
signatures.

\begin{figure*}
  \centering
  \includegraphics[scale=0.37]{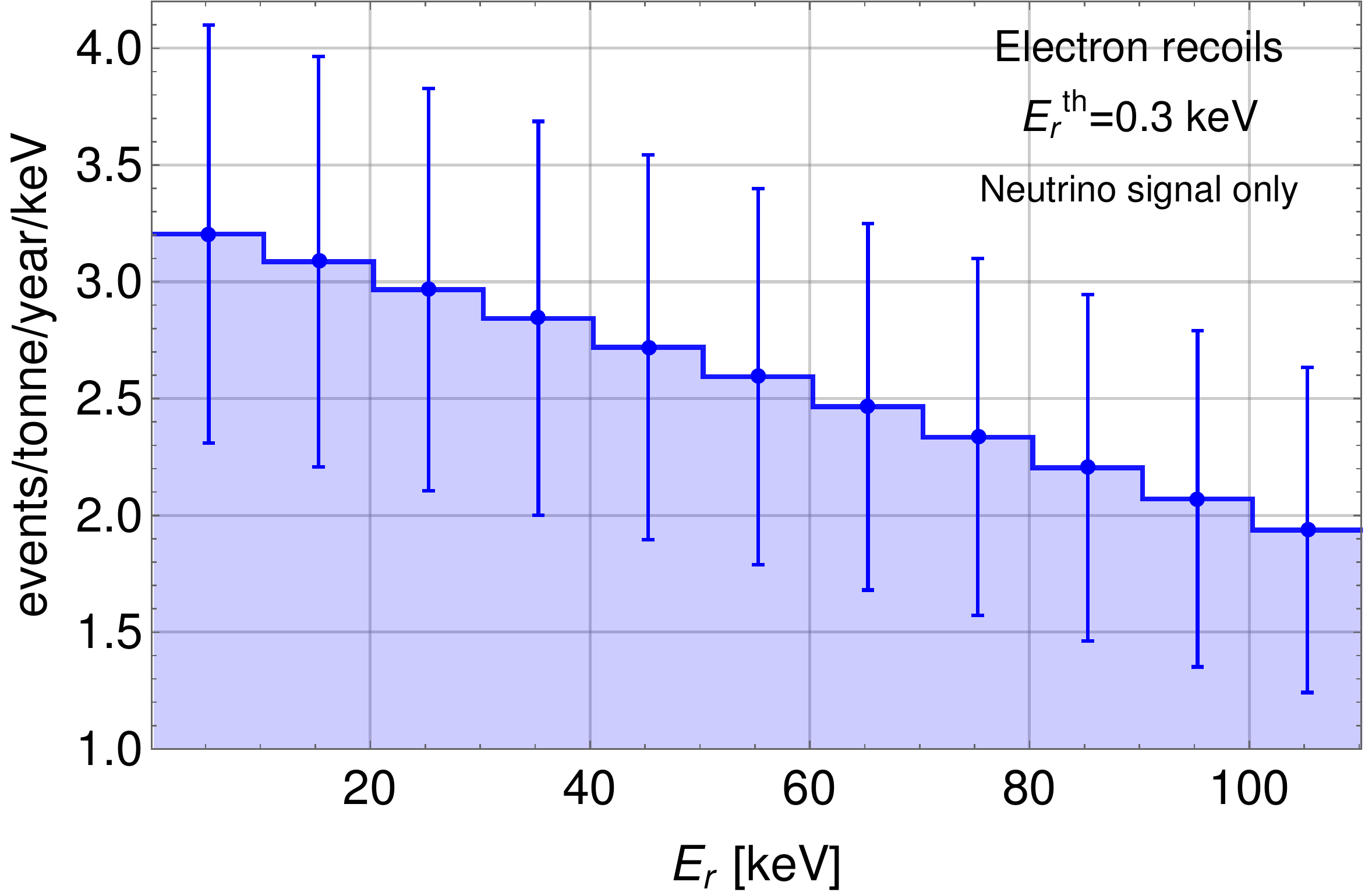}
  \caption{Neutrino-electron signal for a $0.3\,$keV threshold with
    error bars amounting to $68\%$ of the signal rate. The calculation
    assumes 100\% detector efficiency and a one tonne-year
    exposure. Results are rather insensitive to threshold shifts (0.3
    keV to 1 keV).}
  \label{fig:nu-e_signal}
\end{figure*}
To assess the sensitivity of direct detection experiments to neutrino
magnetic moments we use a spectral chi-square test assuming various
detector configurations as follows. One, ten and forty tonne active
volume sizes, 0.3 keV and 1 keV thresholds. For CE$\nu$NS we adopt two
background hypotheses, $68\%$ and $25\%$ of the signal rate. For
$\nu-e$ elastic scattering instead we use the expected backgrounds at
XENON1T, XENONnT and DARWIN reported in
Refs. \cite{Aprile:2020tmw,Aprile:2020vtw,Aalbers:2020gsn}.  They
include material radioactivity, double beta decays of $^{136}$Xe and
$^{124}$Xe decays via double electron capture, among others.  Although
we take these assumptions as representative of XENON1T, XENONnT and
DARWIN, their main motivation is that of comparing the impact of
different active volumes as well as different thresholds and
backgrounds in the reach of direct detection experiments to neutrino
electromagnetic properties.

Under these assumptions we first calculate the signals that define our
toy experiments for both CE$\nu$NS and $\nu-e$ elastic scattering,
considered in all cases as the SM prediction
($\mu_{\nu_\text{eff}}=0$).  For CE$\nu$NS, recoil energies are taken
up to the $^8$B kinematic threshold, $\sim 4.5\,$keV. For $\nu-e$
scattering the pp-induced signal extends up to 264 keV (see
Tab. \ref{tab:kinematic-upper-limits}). However, we consider recoil
energies only up to $105\,$keV, point at which the signal
drops. Covering up to the kinematic limit does not have a substantial
impact in our results.  The resulting toy experiments signals are
shown in Fig.~\ref{fig:signals_CEvNS} for CE$\nu$NS and in
Fig. \ref{fig:nu-e_signal} for $\nu-e$ scattering.  Note that we have
only shown signals at different thresholds for the case of
CE$\nu$NS. We found that for $\nu-e$ scattering, changing the
threshold from $0.3\,$keV to $1\,$keV has a negligible effect, which
means that toy experiments for any of those thresholds produce, in
practice, the same signal. The reason is justified by the fact that
the $0.7\,$keV shift in energy threshold for $\nu-e$ scattering in the
region of interest, reduces the energy range by 0.7\%, while for
CE$\nu$NS by 15\%.

\begin{figure*}[t]
  \centering
  \includegraphics[scale=0.37]{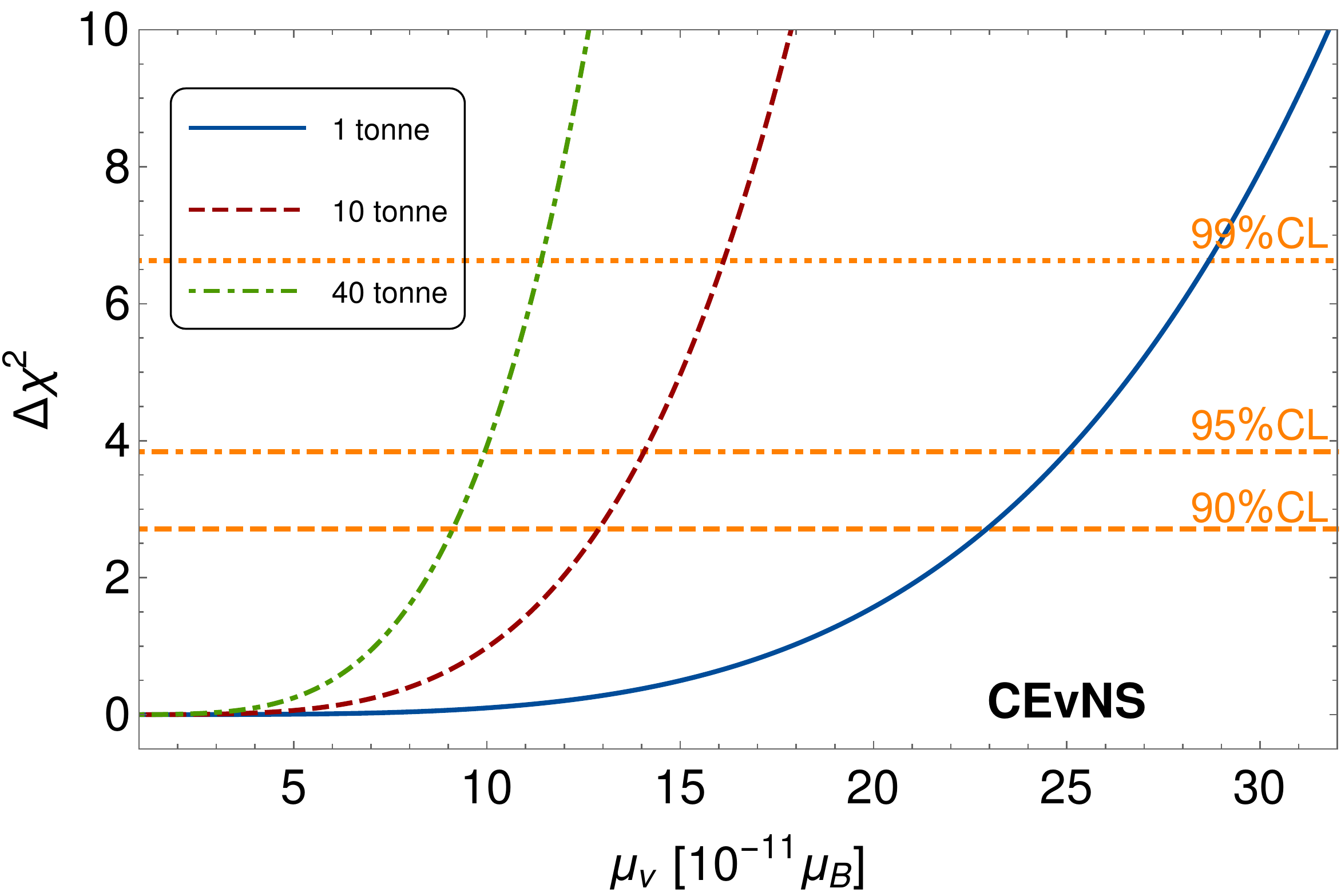}
  \includegraphics[scale=0.37]{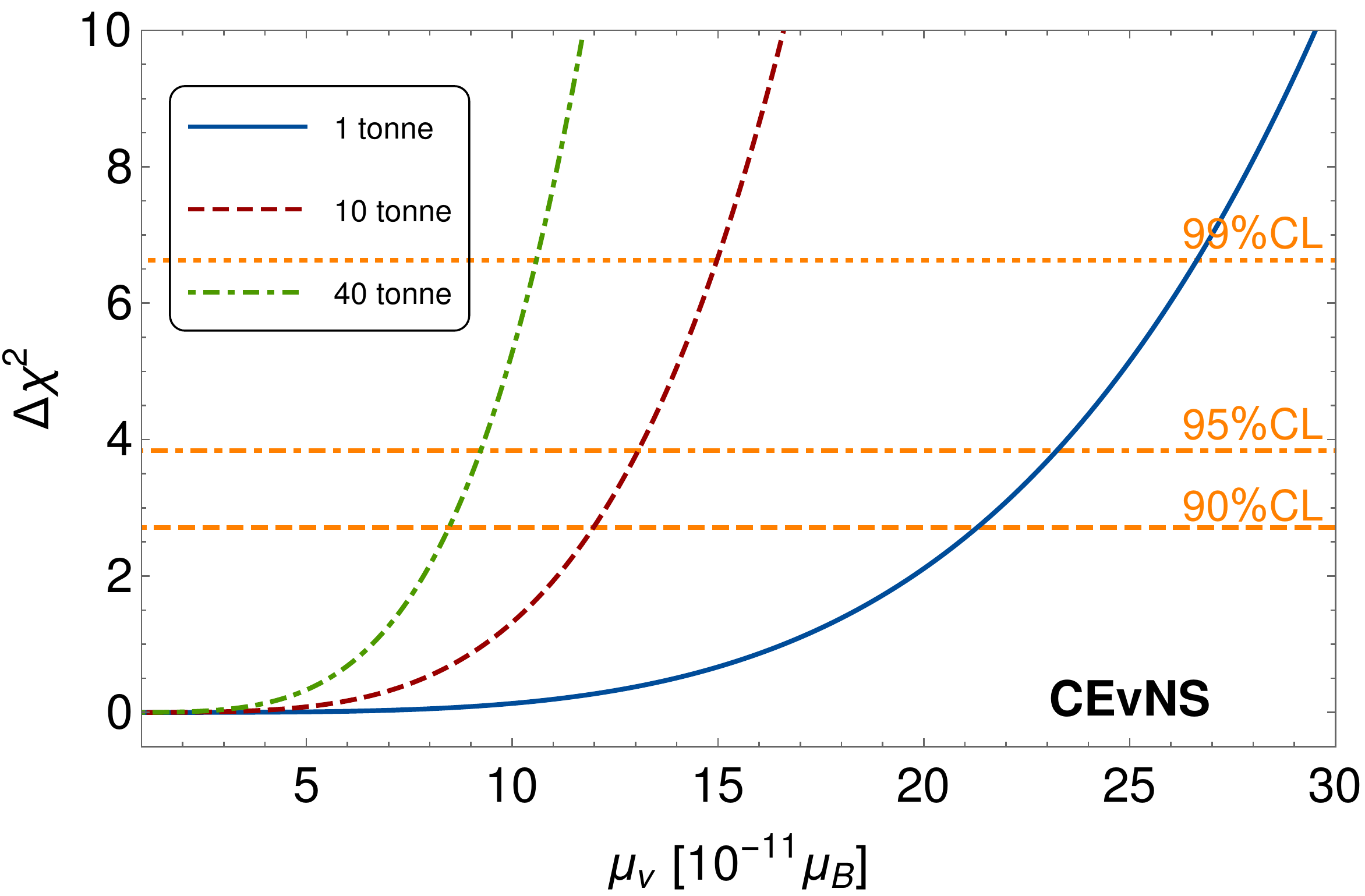}
  \includegraphics[scale=0.37]{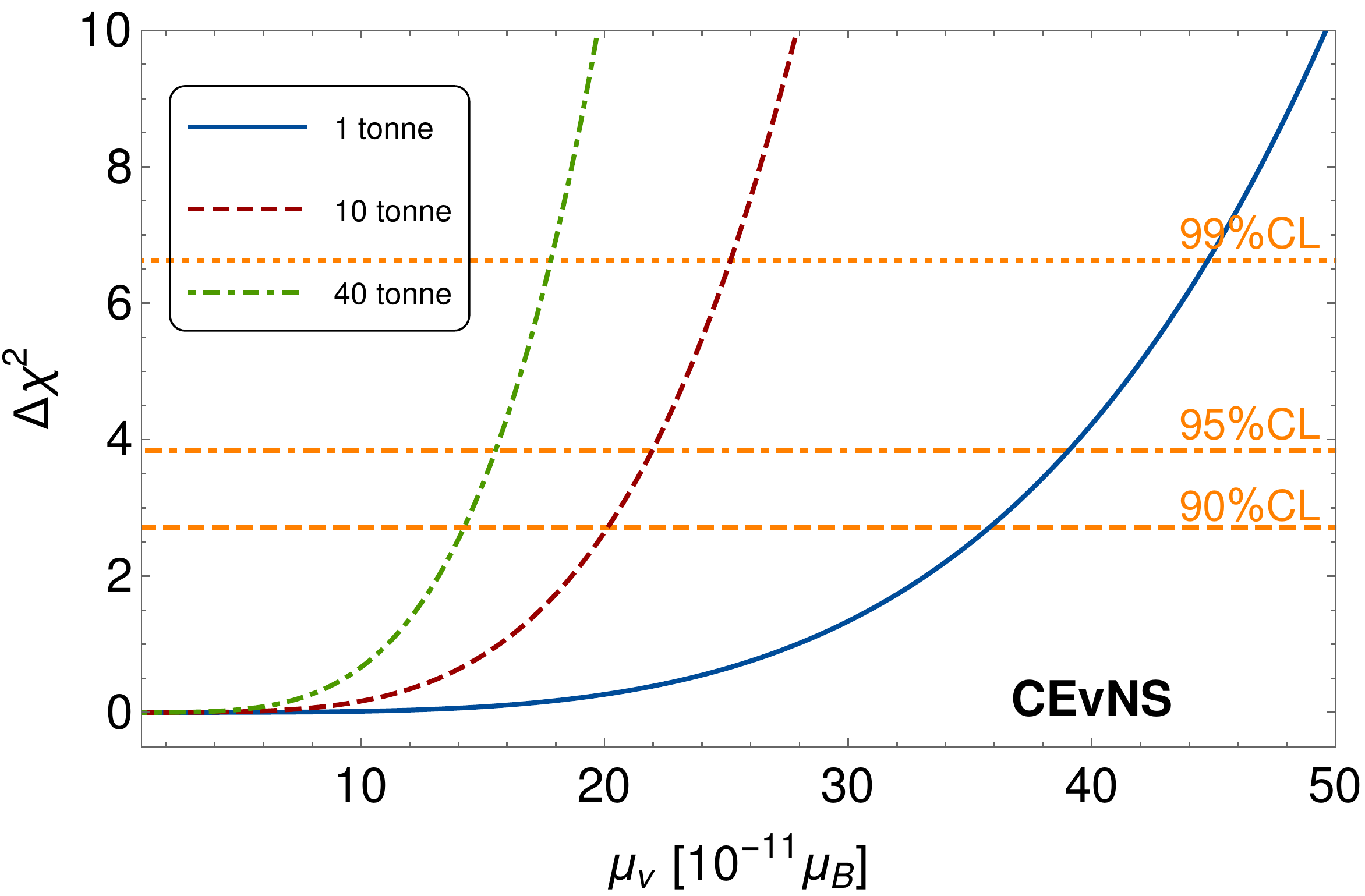}
  \includegraphics[scale=0.37]{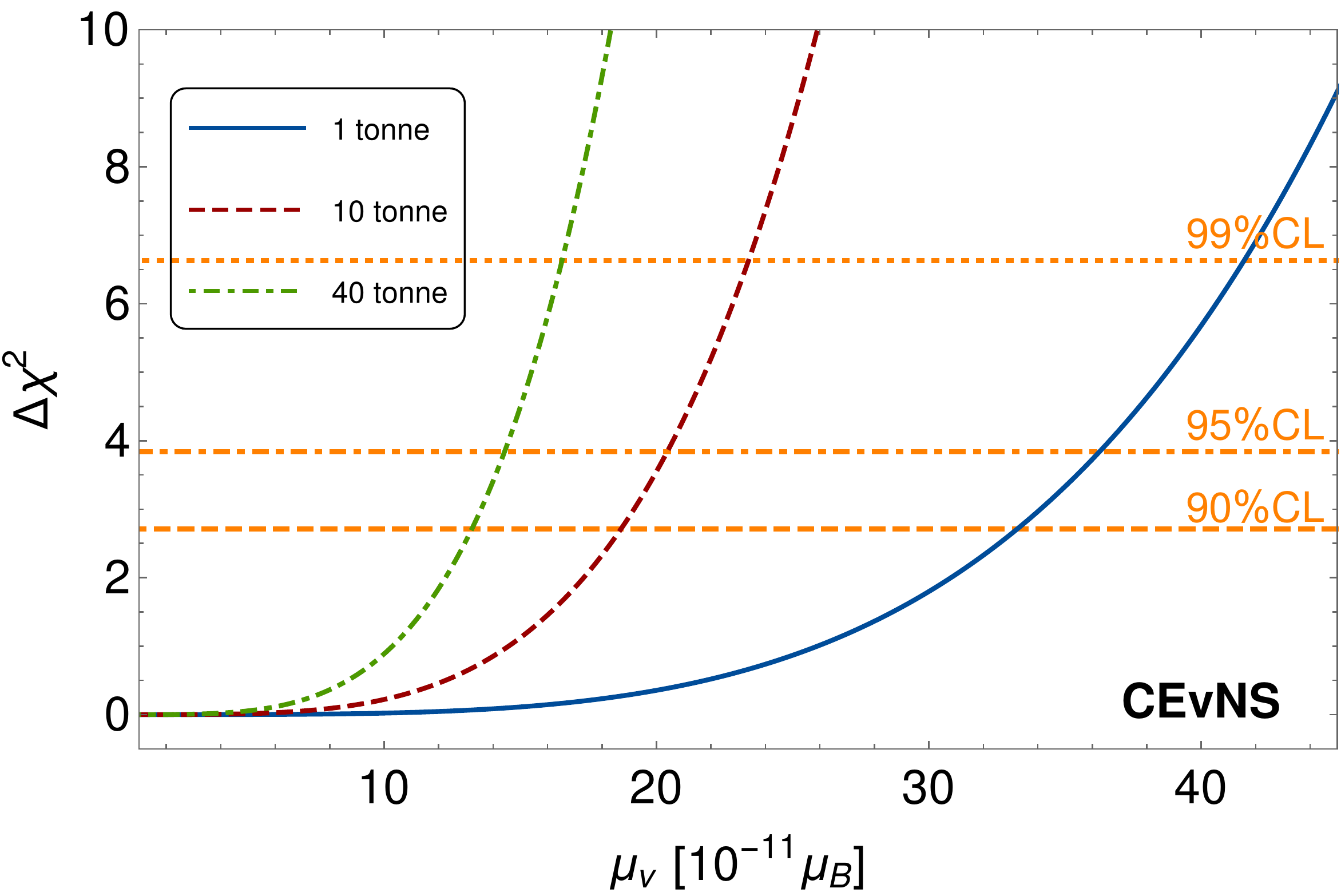}
  \caption{\textbf{Top graphs}: Nuclear recoil sensitivities to
    neutrino magnetic dipole moments in a 1, 10 and 40 tonne active
    volume detectors during a one-year data taking. The result assumes
    a $0.3\,$keV threshold, $100\%$ detector efficiency and
    backgrounds amounting to $68\%$ (left) and $25\%$ (right) of the
    signal rate, Bckg 1 and 2 hypotheses respectively. \textbf{Bottom
      graphs}: Same as top graphs, but assuming a $1\,$keV
    threshold. In contrast to electron recoils, results are rather
    sensitive to threshold performances.}
  \label{fig:chisquare-CEvNS}
\end{figure*}
We define our binned chi-square test according to
\begin{equation}
  \label{eq:chi-square}
  \chi^2=\sum_i\frac{1}{\sigma_i^2}
  \left(
    \left .\frac{dR_i}{dE_r}\right|_\text{SM}
    -\left .\frac{dR_i}{dE_r}\right|_{\mu_{\nu_{\text{eff}}}}
  \right)^2\ ,
\end{equation}
where the recoil spectra in the second term includes both the SM and
neutrino magnetic moment contributions to the signal.  For our
analysis we sample over $\mu_{\nu_\text{eff}}$ from $10^{-9}\,\mu_B$
and as low as $10^{-13}\,\mu_B$ for both CE$\nu$NS and $\nu-e$
scattering channels.  The results of the analysis are shown in
Fig.~\ref{fig:chisquare-CEvNS}, which display $\Delta \chi^2$ versus
$\mu_{\nu_{\text{eff}}}$ (in $\mu_B$ units) calculated for the four
different combinations in Fig.~\ref{fig:signals_CEvNS}.  The results
for different active volume sizes are shown in each plot, proving that
an enhancement from 1 to 40 tonne (XENON1T to DARWIN) will improve the
sensitivity by a factor $\sim 2.5$ at the $90\%$CL.  As can be seen
this sensitivity factor enhancement is independent of threshold and
background conditions.

\begin{figure*}[t]
  \centering
  \includegraphics[scale=0.37]{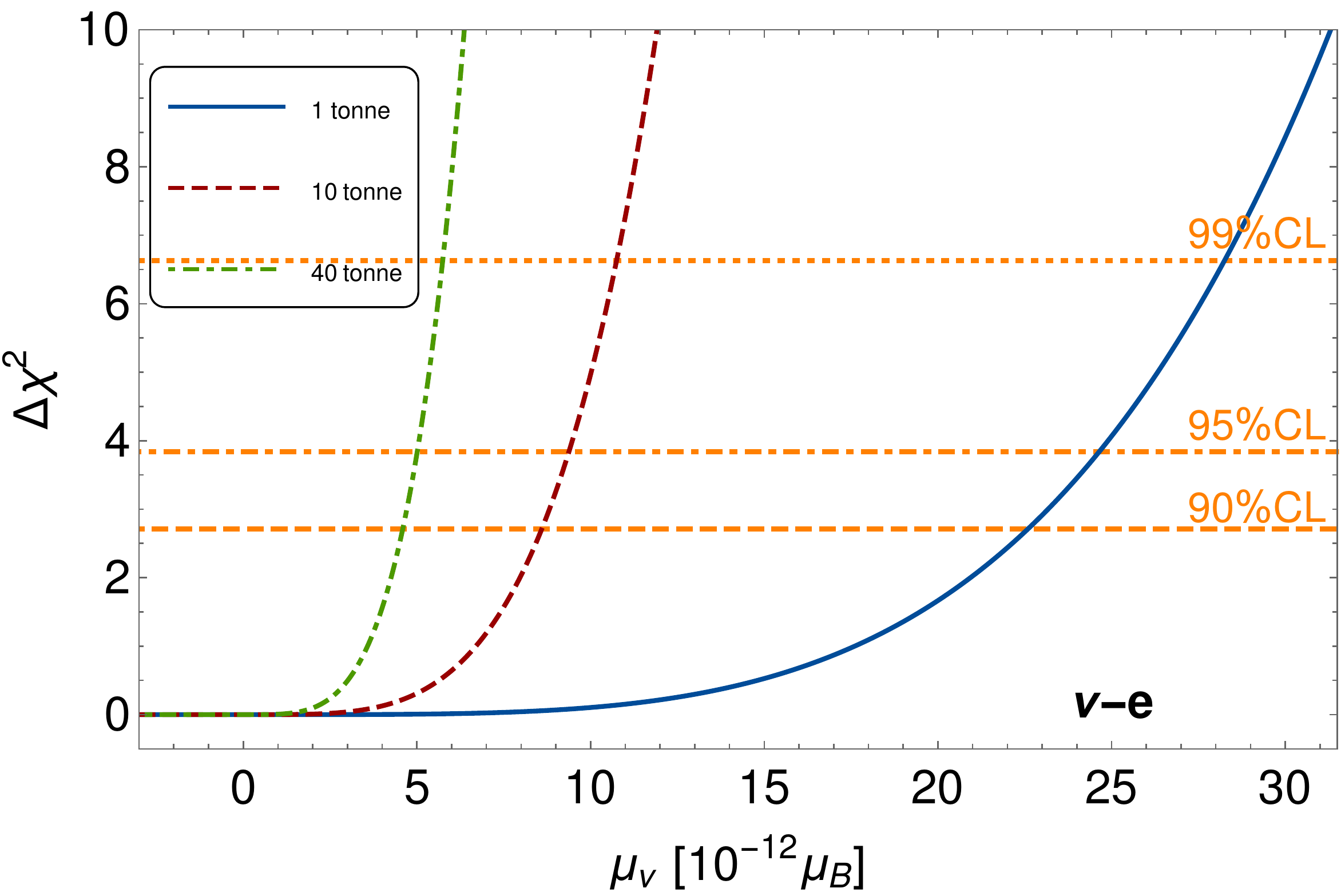}
  \caption{Electron recoil sensitivities to neutrino magnetic dipole
    moments in 1, 10 and 40 tonne active volume detectors during a
    one-year data taking. The result assumes a $0.3\,$keV threshold,
    $100\%$ detector efficiency and backgrounds for XENON1T, XENONnT
    and DARWIN as given in
    Refs. \cite{Aprile:2020tmw,Aprile:2020vtw,Aalbers:2020gsn}. The
    result is rather insensitive to threshold shifts (0.1 keV to 1
    keV).}
  \label{fig:chisquare-nu-e}
\end{figure*}
A direct comparison of the results can be done as follows: panels in
the same row share the same threshold, while those in the same column
share the same background hypothesis.  We can conclude that decreasing
backgrounds may allow a sensitivity enhancement of order $15\%$ at the
$90\%$CL. Clearly, our background hypotheses are somewhat arbitrary.
Changing them will quantitatively affect this conclusion, but the
qualitative feature will be unchanged.  Changing the threshold---as
expected---has a similar impact on sensitivities, they are degraded by
going from 0.3 keV to 1 keV. Overall the best sensitivity is obtained
with a 40-tonne active volume with low background and threshold,
top-right graph where we can see that values as small as
$8.0\times 10^{-11}\mu_B$ at the $90\%$CL can be explored. This result
is remarkable since it shows that if a $0.3\,$keV threshold is
attainable, experiments with characteristics as those of DARWIN will
be able to explore regions in parameter space rather comparable to
those explored by Borexino, TEXONO and GEMMA
\cite{Borexino:2017fbd,Deniz:2009mu,Beda:2010hk} even in nuclear
recoil measurements. Furthermore, it demonstrates that even in its
nuclear recoil data sets DARWIN will be able to test regions close to
those not yet ruled out by astrophysical arguments,
$\mu_\nu|_\text{astro}\lesssim 3\times 10^{-12}\,\mu_B$
\cite{Raffelt:1990pj}. If such threshold is not achievable, and
measurements are ``limited'' to 1 keV threshold instead, with low
backgrounds still sensitivities like those of Borexino, TEXONO and
GEMMA will be within reach in the nuclear recoil channel as the
bottom-right graph in Fig. \ref{fig:chisquare-CEvNS} shows.

For the neutrino-electron scattering case sensitivities are way
better, as expected. Results of our analysis for this process are
shown in Fig.~\ref{fig:chisquare-nu-e}, which displays $\Delta \chi^2$
versus $\mu_{\nu_\text{eff}}$. They have been obtained by using the
toy experiment signal shown in Fig.~\ref{fig:nu-e_signal}, and by
sampling over $\mu_{\nu_\text{eff}}$ in the same range as in the
CE$\nu$NS analysis.  Thanks to the low thresholds and large volume
sizes, sensitivities will outpass those achieved in Borexino, TEXONO
and GEMMA even in the 1-tonne detector case (representative of
XENON1T). At the 90\%CL sensitivities reach values of order
$2.3\times 10^{-11}\mu_B$. Considering the 40-tonne detector instead,
sensitivities improve to values of about $4\times 10^{-12}\mu_B$ at
the 90\%CL.

These values are of the same order and can become more competitive
than those derived from astrophysical arguments, which then brings the
question of the sensitivities that could be reached with other
detector configurations. This question is particularly relevant in the
light of existing theoretical bounds derived using effective theories
or renormalizable models, which lead to values of about
$10^{-14}\,\mu_B$
\cite{Babu:1989px,Barr:1990um,Bell:2005kz,Davidson:2005cs,Bell:2006wi}.
If one takes the bound from astrophysical arguments at face
value\footnote{Note that astrophysical bounds may be subject to
  substantially large uncertainties. So the lower boundary of the
  allowed region should be understood as somewhat fuzzy. This is
  arguably the approach adopted in Ref.~\cite{Aprile:2020tmw} when
  interpreting the electron excess in terms of neutrino magnetic
  moments.}, the region of interest then spans roughly two orders of
magnitude.

\begin{figure*}[t!]
  \centering
  \includegraphics[scale=0.37]{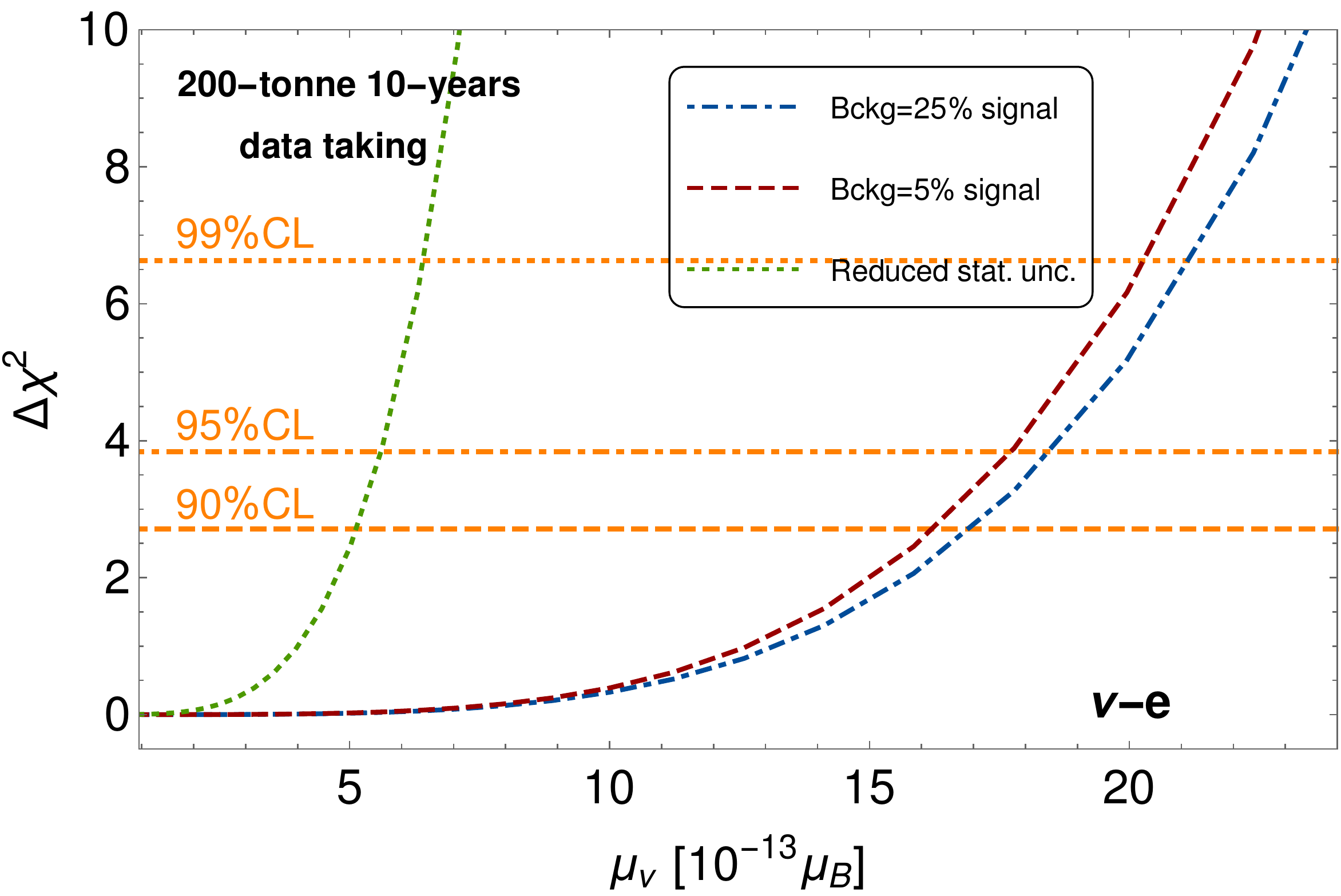}
  \caption{Electron recoil sensitivities to neutrino magnetic dipole
    moments in a 200-tonne liquid xenon detector under the background
    2 and background 3 hypotheses ($25\%$ and $5\%$ of the signal
    rate) assuming 100\% detector efficiency. Results from an analysis
    with a reduced statistical uncertainty are given by the dotted
    green curve. The result is obtained for 10 years of data taking.}
  \label{fig:sensitivities}
\end{figure*}

To determine the degree at which the full region can be covered we
have calculated the sensitivity in electron recoils that could be
achieved in a hypothetical 200-tonne liquid xenon detector under the
most favorable assumptions in ten years of data taking.  We regard
this case as the most optimistic one, and so it fixes the most
ambitious sensitivity one could expect. The result is displayed in
Fig. \ref{fig:sensitivities}, which shows $\Delta\chi^2$ versus
$\mu_{\nu_\text{eff}}$ for the assumed configurations. The result
includes as well an additional background assumption amounting to
$5\%$ of the signal rate, background 3 hypothesis. This result shows
that if low thresholds are achieved in this type of detectors, the
final sensitivity to neutrino magnetic dipole moments will be of order
$10^{-12}\,\mu_B$ with little dependence on background. Under the
background 2 hypothesis the best sensitivity that can be achieved is
about $1.9\times 10^{-12}\,\mu_B$ at the 90\%CL, while with the
background 3 hypothesis this value improves to
$1.8\times 10^{-13}\,\mu_B$ at the 90\%CL. Finally, we have checked
whether reducing statistical uncertainties could allow further
improvements of these sensitivities. Assuming the background 3
hypothesis and reducing $\sigma_a$ in Eq. (\ref{eq:chi-square}) by a
$0.1$ factor we have found that sensitivities could improve close to
values of order $10^{-13}\,\mu_B$.

In summary one can can fairly say that the full region of interest
cannot be covered, but perhaps a reasonable fraction of it
could. Whether this is the case will largely depend on the size of
statistical uncertainties. If they are substantially reduced, these
detectors could eventually test regions of parameter space where
non-zero neutrino magnetic moments could induce sizable signals.
\section{Conclusions}
\label{sec:conclusions}%
With large fiducial volumes DM direct detection experiments are
sensitive to solar neutrinos fluxes. Indeed, the statistics in both
nuclear and electron recoils is expected to be large.  This is the
case, for instance, in XENON1T which has already collected a
substantial number of events in both
channels~\cite{Aprile:2019xxb,Aprile:2020tmw}.  Motivated by the large
statistics expected, in this paper we have studied the sensitivity of
those measurements to neutrino magnetic dipole moments. We have
considered different detector configurations, which although rather
generic are representative of XENON1T, XENONnT and DARWIN. By
generating \textit{toy experiments} signals given by the SM prediction
plus two background hypotheses for nuclear recoils ($68\%$ and $25\%$
of the signal rate) and actual background for electron recoils, we
have done a chi-square test analysis to determine the reach these
detectors would have.

In the case of CE$\nu$NS we have found that sensitivities can be
comparable to those reached by neutrino-electron elastic scattering
dedicated experiments such as TEXONO, Borexino and GEMMA
\cite{Deniz:2009mu,Borexino:2017fbd,Beda:2010hk}. The best sensitivity
can be achieved with the 40-tonne detector, with a 0.3 keV threshold
and low background. In one year of data taking such detector could
explore regions in parameter space down to values of order
$8.0\times 10^{-11}\mu_B$ at the 90\%CL. The 1-tonne detector
operating with the same threshold and low background as well could
achieve values of about $21.5\times 10^{-11}\,\mu_B$ at the 90\%
CL. These sensitivities can be certainly improved with larger data
taking times, but even assuming only one year is already sufficient to
make nuclear recoil measurements competitive with current limits.

Sensitivities with neutrino-electron elastic scattering are way
better. Furthermore, they are rather insensitive to recoil
thresholds. Shifting from $0.3$ keV to $1$ keV changes the
events/tonne/year rate in less than 1\%. In the ideal case of a
40-tonne detector with a 0.3 keV threshold, regions with values as
small as $\sim 4.0\times 10^{-12}\mu_B$ at the 90\%CL could be
explored. This means that using electron recoil measurements these
detectors can explore regions of parameter space not yet ruled out by
astrophysical arguments.  We have found that even the 1-tonne detector
might be able to reach values of order $2.3\times 10^{-11}\mu_B$ at
the 90\%CL in only one year of data taking.  Note that this result is
inline with the neutrino magnetic hypothesis considered by XENON1T in
its electron excess analysis~\cite{Aprile:2020tmw}. These results show
that searches for neutrino magnetic signals are already dominated by
this type of detectors and will keep being so in the future.

Finally, we have quantified the degree at which these detectors could
cover the $10^{-14}\mu_B-10^{-12}\mu_B$ region with increasing data
taking. To do so we calculated sensitivities for a hypothetical
200-tonne detector under two background hypotheses, $25\%$ and $5\%$
of the signal rate and 10 years of data taking. Our findings show that
under these---somewhat extreme---conditions, sensitivities can reach
values of order $1.9\times 10^{-12}\,\mu_B$
($1.8\times 10^{-12}\,\mu_B$) at the 90\%CL for the background 2
(background 3) hypothesis. Covering the full region of interest seems
unlikely, but a reasonable fraction is potentially testable if
statistical uncertainties can be further suppressed. These detectors
thus have a chance to eventually observe neutrino magnetic moment
induced signals.
\section*{Acknowledgments}
We thank Kaixuan Ni for very useful comments on backgrounds as well as
for comments on the manuscript. Jelle Aalbers for providing us details
of the data in Ref. \cite{Aprile:2019xxb} and Dimitris Papoulias for
suggestions.  DAS and RB are supported by the grant ``Unraveling new
physics in the high-intensity and high-energy frontiers'', Fondecyt No
1171136. OGM and GSG have been supported by CONACyT through grant
A1-S-23238.
\bibliography{references}
\end{document}